\documentclass[trans]{IEEEtran}
\usepackage{graphicx}
\usepackage{amsmath}
\usepackage{amssymb}
\usepackage{afterpage}
\usepackage{verbatim}
\usepackage{psfrag}
\usepackage{color}
\usepackage{cite}
\usepackage{lettrine}

\usepackage{setspace}
\usepackage{epsfig}
\usepackage{epstopdf}
\usepackage{footmisc}
\usepackage{fmtcount}
\usepackage{stackrel}
\usepackage{float}
\usepackage{breqn}
\usepackage{enumerate}
\usepackage{subcaption}

\begin{document}

\title {Localization of Energy Harvesting Empowered Underwater Optical Wireless Sensor Networks}

\author{Nasir Saeed,~\IEEEmembership{Senior Member,~IEEE}, Abdulkadir Celik,~\IEEEmembership{Member,~IEEE},  \\  Tareq Y. Al-Naffouri,~\IEEEmembership{Senior Member,~IEEE}, and Mohamed-Slim Alouini,~\IEEEmembership{Fellow,~IEEE}.
\thanks{This work is supported by Office of Sponsored Research (OSR) at King
Abdullah University of Science and Technology (KAUST). 

The authors are with the Department of Electrical Engineering, Computer Electrical and Mathematical Sciences \& Engineering (CEMSE) Division, King Abdullah University of Science and Technology (KAUST), Thuwal, Makkah Province, Kingdom of Saudi Arabia, 23955-6900. A short version of this paper was presented in IEEE ICASSP'18 \cite{Nasir2018limited}.}
}


\maketitle{}
\begin{abstract}
This paper proposes a received signal strength (RSS) based localization framework for energy harvesting underwater optical wireless sensor networks (EH-UOWSNs), where the optical noise sources and channel impairments of seawater pose significant challenges on range estimation. In UOWSNs energy limitation is another major problem due to the limited battery power and difficulty to replace or recharge the battery of an underwater sensor node. In the proposed framework, sensor nodes with insufficient battery harvest ambient energy and start communicating once they have sufficient storage of energy. Network localization is carried out by measuring the RSSs of active nodes, which are modeled based on the underwater optical communication channel characteristics. Thereafter, block kernel matrices are computed for RSS-based range measurements.  Unlike the traditional shortest-path approach, the proposed technique reduces the estimation error of the shortest path for each block kernel matrix.  Once the complete block kernel matrices are available, a closed form localization technique is developed to find the location of every optical sensor node in the network. An analytical expression for the Cramer Rao lower bound (CRLB) is also derived as a benchmark to evaluate the localization performance of the developed technique.  Extensive simulations show that the proposed framework outperforms the well-known network localization techniques.
\end{abstract}

\begin{IEEEkeywords}
 Energy Harvesting, Underwater Optical Wireless Networks, Network Localization, Optical Wireless Communication. 
\end{IEEEkeywords}

\maketitle
\section{Introduction}

\lettrine{U}{nderwater} wireless sensor networks (UWSNs) are enablers of many commercial, scientific and military underwater applications, including instrument monitoring, climate recording,  prediction of natural disasters, exploration for the oil industry, search \& rescue missions, control of the autonomous underwater vehicles (AUVs), and marine life study \cite{AKYILDIZ2005257, Vasilescu:2005}. The need for high quality of service communication necessitates high data rate, low latency, and long-range networking solutions poses a daunting challenge because of the highly attenuating medium of seawater for most electromagnetic frequencies. 

Today's UWSNs are mostly based on acoustic communication systems that suffer from unique aquatic conditions and display severe attenuation characteristics, frequency dispersion, multipath fading, and limited bandwidth. In addition, the signal propagation delays and variable speed of sound create a set of unique challenges. Therefore, underwater acoustic communication has low achievable rates (10-100 kbps), due to the limited bandwidth and high latency caused by the low propagation speed of acoustic waves (1500 m/s)  \cite{7593257}. On the other hand, underwater optical wireless communication (UOWC) has the advantages of having a higher bandwidth, lower latency, and enhanced security \cite{Kaushal2016underwater}. Nevertheless,  UOWC has a very limited range attainability (10-100 m) due to absorption, scattering, and turbulence impairments of seawater. It is also susceptible to many noise sources such as sunlight, background, thermal, and dark current noises \cite{Akhoundi2016cellular}.  Accordingly, underwater sensor nodes communicating through UOWC require the development of a densely deployed multi-hop underwater optical wireless sensor network (UOWSN). The potential applications of densely deployed UOWSNs include, but are not limited to, the monitoring of underwater oil and gas pipelines, security and surveillance, ocean exploration, and the monitoring of underwater habitat.

Furthermore, underwater sensor nodes have limited energy resources, which has a substantial impact on the lifetime of the network. Taking the engineering hardship and monetary cost of battery replacement into account, it is clear that an energy self-sufficient UWSN is essential to maximize the lifetime of the network. To do so, we consider energy harvesting as a promising potential solution to collect energy from  ambient sources in the aquatic environment, and then to be stored in an energy buffer \cite{Nasir2018}. As reported in  \cite{EHsurvey}, ongoing research efforts on terrestrial communications have shown that energy harvesting plays a significant role in improving the performance.  However, most of the energy-harvesting techniques are designed for outdoor environments and not applicable to an aquatic environment. Moreover, albeit the notable research body on designing different protocols for underwater communication networks, no significant research is carried out on the energy harvesting methods for UOWSNs.

As the gathered data are useful only if they refer to a particular position of the sensor node, the localization of nodes in UOWSN is of utmost importance. Network localization is especially useful for  applications such as target detection, intruder detection, routing protocols, and data tagging. Hence, a number of acoustic underwater sensor network localization techniques have been proposed in the past \cite{1acoustic,2acoustic, 3acoustic,  4acoustic, 5acoustic, 6acoustic, kantarci2011survey}. The performance of these localization techniques mainly relies on the initial reference position, number of sensor nodes, ranging technique, number of anchors and the position of the anchors in the network \cite{TAN20111663}. However, the aforementioned optical communication challenges and energy constraints do not allow the use of existing acoustic localization techniques for underwater sensor nodes. In \cite{Nasir2018icc}, the connectivity analysis of UOWSNs is investigated where an analytical expression is derived which shows that the probability of a network to be connected directly depends on the number of nodes in the network, the beamwidth of the transmitted signal, and transmission range of the nodes. The UOWSN localization technique is addressed in \cite{Akhoundi2017underwater} where the time of arrival (ToA) and RSS methods are investigated in the case of an optical code-division multiple access network. Consequently, an RSS-based localization technique is developed in \cite{Saeed2018tcom}, which takes into consideration the outliers in ranging and optimizes the location of anchor nodes to improve the localization accuracy in UOWSNs. ToA-based localization schemes can provide a higher precision, heavily relying on synchronization and additional clocks to measure the time of transmission, yielding to extra hardware complexity and cost. However, RSS-based localization schemes are generally preferred because of their simplicity and cost efficiency.
In this paper, we investigate a RSS-based localization technique for energy harvesting underwater optical sensor networks (EH-UOWSNs). To the best of our knowledge, this has never been investigated in the past.

\subsection{Main contributions}
The main contributions of this paper can be summarized as follows:
\begin{itemize}
\item We consider an energy-harvesting UOWSN setup where nodes harvest and store ambient renewable energy sources and individually optimize their duty cycles to achieve maximum throughput. We have shown that energy arrival rates play a critical role in network connectivity since it primarily determines the number of active sensors and thus the degree of network connectivity. Furthermore, it has been shown that higher energy arrival rates and better network connectivity are two main factors that improve the performance of the network localization. Numerical results show that the harvesting energy from multiple underwater sources improves the connectivity of the network, which, in turn, improves the accuracy of the proposed network localization scheme. 

\item 
A closed-form localization technique is developed to estimate the location of each optical sensor node in UOWSN. The proposed algorithm is a modified iterative majorization formulation that improves the accuracy  of the localization by directly including the anchor's location in the estimation process. Additionally, the proposed approach does not require any extra transformation, contrary to its counterparts. The proposed solution accurately minimizes the error function by partitioning the kernel matrix into smaller block matrices. In order to obtain a better approximation of the missing distance measurements, a novel matrix completion strategy is also developed to complete the missing elements in the block matrices. Finally, an analytical expression for Cramer Rao lower bound (CRLB) is derived for the proposed UOWSN localization technique. Simulation results show that the root mean squared positioning error (RMSPE) of the proposed technique is more robust and accurate than the well-known network localization techniques such as Isomap \cite{Kashniyal2017}, multidimensional scaling (MDS) \cite{Shang, nasir2015}   and local linear embedding (LLE) \cite{Roweis2323}.

\end{itemize}

\subsection{Paper Organization}
The rest of the paper is organized as follows: Section \ref{sec:harvesting} introduces the energy harvesting framework for UOWSNs and discuss the transmission strategies based on the energy arrival-rate. In Section \ref{sec:proposed}, we define the system model, formulate the problem, and propose the localization method for UOWSNs. Section \ref{sec:analysis} presents the performance analysis, in terms of CRLB, for the proposed technique. In section \ref{sec:results}, simulations are conducted for evaluating the performance of the proposed technique; section \ref{sec:conc} concludes the proposed work. 

\section{Impacts of Energy Harvesting on Network Connectivity and Localization}\label{sec:harvesting}
Replacing or recharging the battery of sensors is a challenging engineering task, especially in deep oceanic waters; this is why energy self-sustainability plays a crucial role in maximizing the network lifetime of UOWSNs. The set of active sensors that affects the degree of connectivity is mostly determined by the power availability. Therefore, it makes sense that the connectivity has a great impact on the localization performance, as having more active nodes provide more distance measurements that naturally improve the localization accuracy. Harvesting energy from ambient renewable energy sources is a potential promising solution to provide an energy-self-sufficient operation of UOWSNs, and to improve the location accuracy. 

Unlike traditional battery powered networks, energy harvesting requires that the stored energy, at a certain time, is greater than, or equal to, the consumed energy up to this time. This phenomenon, referred to as the, \textit{energy causality constraint} (ECC),  can be defined, for each sensor node $i \in [1, N]$, as
\begin{align}
\label{eq:ECC}
\nonumber & B_i^0 + \eta_i^s \int_0^{\mathcal{T}} \left[P^h_i(t) -P^c_i(t) \right]^+dt \geq \\
\int_0^{\mathcal{T}}&  \left[P^c_i(t) -P^h_i(t) \right]^+dt + \int_0^{\mathcal{T}} P_\ell \: dt, \mathcal{T} \in [0, \infty), 
\end{align}
where $B_0^i$ is the initial energy buffer state, $ \eta_i^s$ is the storing efficiency of the energy buffer, $P^h_i(t)$ is the harvested energy (possibly from several harvester types) at time $t$, $P^c_i(t)$ is the consumed energy at time $t$, $P_\ell$ is the leak power of the energy buffer, and $\left[ x \right]^+ \triangleq \max(0,x)$. Therefore, it is necessary to decide on a transmission policy based on an optimal duty-cycling for active and passive time portions of sensors. Thereafter, we optimize duty-cycles subject to a minimum duty cycle, ECCs, and buffer capacity.

Energy availability and network connectivity are two coupled factors that play an important role in the performance of the network localization. On one hand, higher energy arrivals to an energy harvesting network improves the network connectivity by  allowing more nodes to be active  and expanding the coverage region of the active nodes. On the other hand, a better network connectivity substantially improves the localization performance, as having more pairwise range measurements intuitively reduces the localization errors. Moreover, leveraging a more accurate location information enables precise pointing, acquisition, and tracking (PAT) mechanisms of optical transceivers to sustain reliable single-hop links. Network connectivity and localization can also be considerably improved by multi-hop communication over these reliable single-hop links, which can be enabled by effective geographic routing algorithms relying on the precise node locations. This means that the energy availability sets a reciprocal relationship between the degree of network connectivity and the performance of the localization, with each susceptible to be influenced by the other.

In this respect, we consider a network of $N$ energy-harvesting nodes that periodically sense and opportunistically transmit a certain number of packets, based on energy availability. The degree of connectivity for power-constrained underwater wireless networks typically relies on low-cost metrics where the objective is to choose a minimum energy-consumption path, from the source to the destination node.  Also, the range between any two neighbor nodes is a function of the transmission power. Therefore, as the amount of energy entering into the network increases, it improves the range $r$ of the nodes which in turn improves the connectivity of the network. The probability of having a connected network is equal to 1 if $r \geq \sqrt{c\log(N_a)/N_a}$, where $N_a$ is the total number of active nodes and $c$ is constant \cite{Betstetter2004}. Thus, maximizing the total number of active nodes for a given communication range can improve the connectivity of the network and its localization.

\begin{figure}
\begin{center}
\includegraphics[width=0.9\columnwidth]{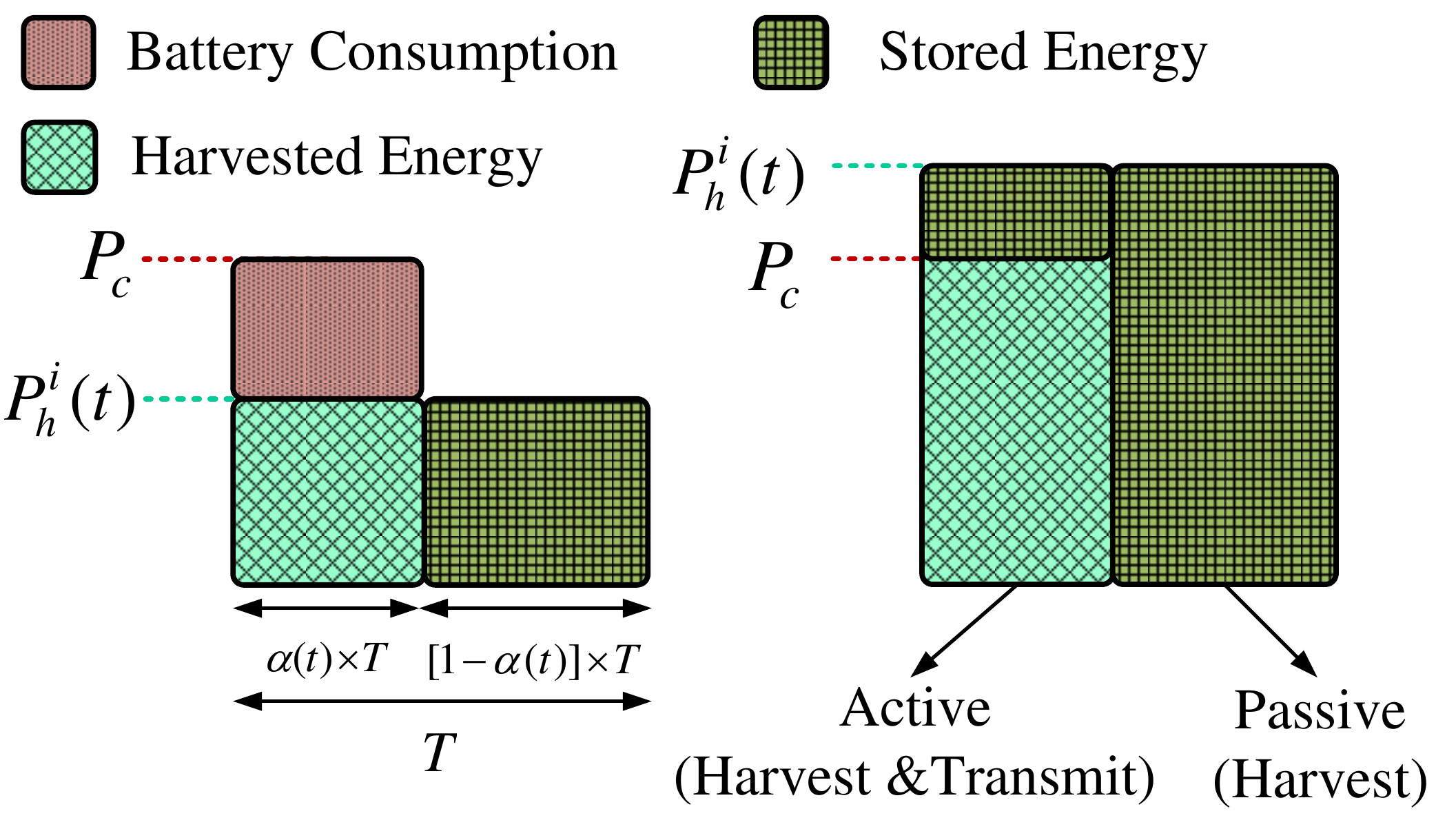}
 \caption{Illustration of the considered time-slotted operation of sensor nodes.}
 \label{fig:duty}
\end{center}
\end{figure} 
Accordingly, we optimize the duty cycle of sensors in order to increase the number of active nodes for a better network connectivity. Each sensor operates on a time-slotted operation (Fig.1) in which the duration of the $t^{th}$ time slot, $T$, is split into active and passive states of duration $\alpha(t) T$ and $[1-\alpha(t)] T$, respectively. In the active state, sensors execute harvesting and transmission momentarily. If the energy-arrival rate to node $i$, at time slot $t$, $P_h^i(t)$\footnote{We assume that energy arrival rate to node $i$ at time slot $t$, $P_h^i(t)$, is perfectly estimated based on historical data and does not change during a time slot duration.}, is less than the consumption power $P_c>P_h^i(t)$ (as in the left hand of Fig. \ref{fig:duty}), power gap is filled by the energy stored in the battery.  However, if $P_c \leq P_h^i(t)$,  the excess harvested energy is stored in the battery (as in the right hand of Fig. \ref{fig:duty}). Sensors switch into sleep mode and only harvest energy in the passive mode where the sleeping power is considered to be negligible.

Accordingly, each node optimizes its duty cycle in order to maximize the throughput over $\mathcal{T}$ time slots, as follows
\begin{equation*}
\hspace*{0pt}
 \begin{aligned}
 & \hspace*{30pt}   \underset{\pmb{\alpha},\pmb{B}}{\max}
& & \hspace*{3 pt}  \frac{1}{\mathcal{T}}\sum_{t=1}^{\mathcal{T} } \alpha(t)\\
& \hspace*{0pt} \mbox{$\mathrm{C}_o^1$:}\hspace*{20pt} \text{s.t.}
&& B(0) = B_0\\ 
 &
\hspace*{0 pt}\mbox{$\mathrm{C}_o^2$: } & & B(t)=B(t-1)+\eta_s T \alpha(t) \left[P_h(t)-P_c \right]^+  + \\
&
& & \hspace*{-15 pt} \eta_s T P_h(t)[1-\alpha(t)] - T \alpha(t) \left[P_c -P_h(t)\right]^+ - T P_\ell,  \hspace{70pt}  1 \leq t \leq \mathcal{T} \\
& 
\hspace*{0 pt}\mbox{$\mathrm{C}_o^3$: } & & 0 \leq B(t) \leq B_{\max}, \hspace{168pt}  1 \leq t \leq \mathcal{T}\\
    &
  \hspace*{0 pt}\mbox{$\mathrm{C}_o^4$: } & & 0 \leq \alpha(t) \leq 1, \hspace{188pt} 1 \leq t \leq \mathcal{T}
\end{aligned}
\end{equation*}
where $\pmb{\alpha}=[\alpha(1), \ldots, \alpha(\mathcal{T})]$, $\pmb{b}=[b(1), \ldots, b(\mathcal{T})]$, $b_0$ is the initial battery energy, and $b_{\max}$ is the battery capacity. 
$\mathrm{C}_o^1$ introduces the initial battery level. The evolution of the state of energy of the battery level is expressed in $\mathrm{C}_o^2$ where the first term is the battery level of the previous time slot, the second term is the energy stored in the active state, the third term is the energy stored in the passive state, the fourth term is the energy consumption, and the fifth term is the battery leakage during a time slot duration. Finally, variable domains for battery levels and duty-cycle are given in $\mathrm{C}_o^3$ and $\mathrm{C}_o^4$, respectively. This problem is apparently a linear programming problem that can be solved rapidly, using the interior-point method. As per the optimal duty-cycles, $\pmb{\alpha}^*$, nodes either go active to transmit and harvest,  if $\alpha(t)>\beta$, or passive to harvest only if $\alpha(t)<\beta$, where $\beta$ is a threshold level determined based on minimum transmission time to transmit a certain number of packet at a given data rate. Notice that duty-cycle optimization intuitively increases the number of time slots where a sensor node can be active and thus improving the total number of active nodes, $N_a$.

\section{Problem Statement and Proposed Localization Method}\label{sec:proposed}
\label{sec:formula}
In this section, we first define the system model and formulate the UOWSN localization problem.  Next, we introduce the proposed localization method; and lastly  the energy cost of computing the pairwise measurements is analyzed.
\subsection{System Model and Problem Statement}   
\begin{figure}
\begin{center}
    \includegraphics[width=0.9\columnwidth]{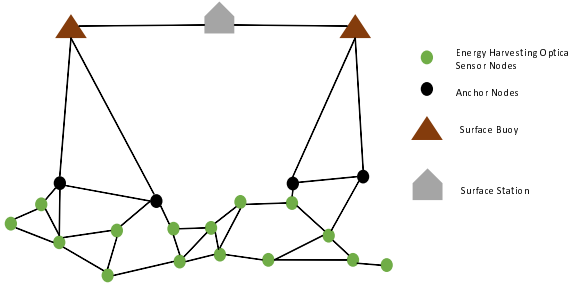}
    \caption{System setup for an EH-UOWSN.}
    \label{fig:setup}
    \end{center}
    \end{figure}

Consider an EH-UOWSN consisting of  $N$  energy-harvesting optical sensor nodes, $M$  anchors, surface buoys and a surface station in a $\rho$-dimensional Euclidean space, as shown in Fig.~\ref{fig:setup}. Each node performs optical ranging and compute the RSS values. The RSS values are forwarded to the surface station by the surface buoys.   The surface station collects all the ranging measurements and estimate the location of each node.

The problem of energy harvesting empowered UOWSNs localization is to determine the position of underwater optical sensor nodes in a $\rho$-dimensional Euclidean space given that some of the pairwise noisy RSS-based optical range measurements and few anchors are available in the network. Additionally, we consider that the energy-harvesting capability of the sensor nodes improves the connectivity of the network. Based on the aforementioned considerations, the network is represented as a weighted graph  $G_{\rho,K}(\boldsymbol{Y},\boldsymbol{\Lambda}, \boldsymbol{D})$ where $\boldsymbol{Y} = \{\boldsymbol{y}_1,...,\boldsymbol{y}_{M},\boldsymbol{y}_{M+1},...,\boldsymbol{y}_{K}\}$ are the coordinates of vertices/sensors, $K=M+N_a$, $\boldsymbol{\Lambda}=\{\Lambda_{i,j}\}_{i,j=1,i\neq j}^{K}$ are edge weights characterized by the link quality, and $\boldsymbol{D} = \{\tilde{d}_{i,j}\}_{i,j=1,i\neq j}^{K}$ are the estimations of the associated distances. The initial number of active and passive sensor nodes are denoted by $N_a$  and $N_p$ (i.e., $N=N_a+N_p$), respectively.

From the above description, only single-hop noisy RSS-based optical range measurements are available; therefore, it is required to estimate the missing multihop pairwise distances in a graph $G_{\rho,K}(\boldsymbol{Y},\boldsymbol{\Lambda}, \boldsymbol{D})$. After estimating the missing multihop pairwise distances, an estimation of the final position is required for all  nodes in the network, by using the position of the anchor nodes.

\subsection{Ranging in UOWSNs}
Optical light passing through the aquatic medium suffers from widening and attenuation in angular, temporal and spatial domains \cite{Akhoundi2017underwater}. The widening and attenuation of the underwater optical signals depend on the wavelength. The extinction coefficient is modeled as a combination of the absorption coefficient $a(\lambda)$ and scattering coefficient $s(\lambda)$, as follows  \cite{shifrin1998}
\begin{equation}
e(\lambda)=a(\lambda)+s(\lambda).
\end{equation} 
Accordingly, the channel gain driven by the absorption and scattering effects of the aquatic medium can be given as 
\begin{equation}
L_{ij} = \exp^{-e(\lambda)d_{ij}}
\end{equation}
where $d_{ij} =\sqrt{(\boldsymbol{y}_i-\boldsymbol{y}_j)(\boldsymbol{y}_i-\boldsymbol{y}_j)^T}$ is the Euclidean distance between nodes $i$ and $j$. Assuming a line-of-sight link from node $i$ to node $j$, the received signal strength (RSS) at sensor node $j$ is given by \cite{Arnon:09}
\begin{equation}\label{eq: pr}
P_{r_{ij}}= P_{t_i}\eta_{i}\eta_jL_{ij}\frac{A_j\cos\theta}{2\pi d_{ij}^2(1-
\cos\theta_0)},
\end{equation}
where $P_{t_i}$ is the optical power transmitted by node $i$, $\eta_{i}$ and $\eta_j$ are the optical efficiencies of node $i$ and $j$, respectively, $A_j$ is the aperture area of node $j$, $\theta$ is the angle between the transmitter trajectory node $i$ and node $j$ receiver, and $\theta_0$ is the divergence angle of the transmitted signal. A transfer function $f \colon P_{r_{ij}} \to d_{ij}$ is necessary to obtain a range from RSS measurements, which can be derived from \eqref{eq: pr}, as follows
\begin{equation}\label{eq: dij}
d_{ij} \triangleq  f(P_{r_{ij}}) =\frac{2}{e(\lambda)} W_0 \left(\frac{e(\lambda)}{2}\sqrt{\frac{ P_{t_i}\eta_{i}\eta_j A_j\cos\theta}{P_{r_j} 2\pi (1-\cos\theta_0) }}\right)
\end{equation}
where $W_o(\cdot)$ is the real part of Lambert-W function  \cite{Corless1996}. In \cite{Akhoundi2017underwater}, the same path loss model is considered for underwater optical wireless positioning where the RSS based distances are estimated by using Monte Carlo simulations. Authors of  \cite{Pelka2017} conducted experiments to justify the model presented above, at different wavelengths with different absorption and scattering coefficients. Notice that the received power expression in \eqref{eq: pr} merely considers attenuation, scattering, and geometrical losses. Nonetheless, the photo-current observed at the output of the photo-detector is subject to thermal and optical noise sources. Accordingly, we consider three primary noise sources \cite{Lapidoth2009}: 1) The thermal noise of the receiver that is well-modeled by a Gaussian distribution, 2) the relative-intensity noise that models random intensity fluctuations and can also be considered to be Gaussian, and 3) the shot noise caused by optical filtering to minimize the ambient/background light.  The shot noise has a considerable impact at large intensities where its distribution also tends to be Gaussian \cite{Kahn1997}, whereas the thermal noise is the limiting factor at low intensity. The superposition of these main noise sources can be well-modeled by independent and memoryless additive Gaussian noise \cite{Lapidoth2009} as follows
\begin{equation}\label{eq: pr2}
\tilde{P}_{r_{ij}} = P_{r_{ij}}+n_{ij},
\end{equation}
where $n_{ij}$ is modeled as a zero mean Gaussian random variable $n_{ij} \sim \mathcal{N}\left(0,\sigma_{ij}^2 \right)$ with variance $\sigma_{ij}^2$. Based on \eqref{eq: pr2}, the range measurement can be obtained as $\tilde{d}_{ij}=f(\tilde{P}_{r_{ij}})$ which is erroneous and adds an extra uncertainty to any localization algorithm. 
Building upon these erroneous range measurements, the remainder of this section presents the proposed localization algorithm.

\subsection{Proposed Network Localization Method}
To model the senor's area of coverage, we consider a circular-noise disk model, that is, $i$th sensor will communicate directly with any other sensor in its transmission capability $r_i$. Notice that $i^{th}$ sensor can also communicate with any sensor located outside of its transmission capability in a multi-hop transmission fashion.  
It is important to note that having multiple RSS measurements can characterize the temporal nature of the estimation noise, and thus improve the localization via more accurate range estimations\footnote{Multiple measurements can be taken within the active period of duty cycle while nodes transmit data packages.}. However, increasing the number of measurements has a higher cost, as the one-time ranging between a pair of nodes requires $O(K^2)$ measurements for $K$ number of nodes. 
The error function between the estimated distances and the actual Euclidean distances is defined as
\begin{equation}\label{Eq: S}
\mathrm{S}(\boldsymbol{Y})=\sum_{i<j<K}\Lambda_{ij}\left(\tilde{d}_{ij}-d_{ij}(\boldsymbol{Y})\right)^2,
\end{equation}
where $\Lambda_{ij}$ represents the quality of the link between node $i$ and $j$. Quality of the link between any two nodes $i$ and $j$ is defined based on  the range error variance, i.e., $\Lambda_{ij}=\frac{1}{\sigma_{ij}^2}$, where the noisy measurements are down-weighted by the large noise variance and vice-versa. The error function in (\ref{Eq: S}) is  expanded as
\begin{eqnarray}\label{eq: stress}
\mathrm{S}(\boldsymbol{Y})&=&\sum_{i<j}\Lambda_{ij}\tilde{d}_{ij}^2+\sum_{i<j}\Lambda_{ij}d_{ij}^2(\boldsymbol{Y})\nonumber \\ & &-2\sum_{i<j}\Lambda_{ij}\tilde{d}_{ij}d_{ij}(\boldsymbol{Y}).
\end{eqnarray}
The error function in \eqref{eq: stress} can be solved by using an iterative majorization approach \cite{Leeuw2009}. The key idea behind the majorization is to replace iteratively the actual complicated function $\mathrm{S}(\boldsymbol{Y})$ by an auxiliary function that is simpler and always has smaller values than the actual function \cite{Leeuw2009}.  The first term in \eqref{eq: stress} is constant and only depend on $\Lambda_{ij}$ and $\tilde{d}_{ij}$ , therefore it is omitted from the iterative majorization approach. The second term is quadratic in $\boldsymbol{Y}$ and is the weighted sum of $d_{ij}^2(\boldsymbol{Y})$, and the last term is a weighted sum of distances. To obtain the majorization of $\mathrm{S}(\boldsymbol{Y})$, we use the Cauchy-Schwarz inequality, i.e.,
\begin{eqnarray}
d_{ij}(\boldsymbol{Y})&=&\parallel\boldsymbol{y_i}-\boldsymbol{y_j}\parallel = \parallel\boldsymbol{y_i}-\boldsymbol{y_j}\parallel\frac{\parallel\boldsymbol{v_i}-\boldsymbol{v_j}\parallel}{\parallel\boldsymbol{v_i}-\boldsymbol{v_j}\parallel}\geq  \nonumber \\ & & \frac{(\boldsymbol{y}_i-\boldsymbol{y}_j)^T (\boldsymbol{\Lambda}_i-\boldsymbol{\Lambda}_j)}{\parallel\boldsymbol{v_i}-\boldsymbol{v_j}\parallel},
\end{eqnarray}
where $\boldsymbol{\Lambda}=\{\boldsymbol{\Lambda}_1,...,\boldsymbol{\Lambda}_{K}\}^T \in \mathbb{R}^{K \times \rho}$ and $\boldsymbol{v_i}$ is the column vectors of $\boldsymbol{V}=\mathbf{\Lambda} \circ \mathbf{A}  $ where $\circ$ is the dot product and $\mathbf{A} \in \{0,1\}^{K \times K}$ is the adjacency matrix representing the connectivity. Therefore, the last term is bounded by
\begin{equation}
\label{eq: CSI}
\sum_{i<j}\Lambda_{ij}\tilde{d}_{ij}d_{ij}(\boldsymbol{Y}) \geq \sum_{i<j}\Lambda_{ij}\tilde{d}_{ij}\frac{(\boldsymbol{y}_i-\boldsymbol{y}_j)^T (\boldsymbol{\Lambda}_i-\boldsymbol{\Lambda}_j)}{\parallel\boldsymbol{v_i}-\boldsymbol{v_j}\parallel}.
\end{equation}
Following from \eqref{eq: stress}-\eqref{eq: CSI}, the error function in \eqref{Eq: S} is bounded by $\mathrm{G}(\boldsymbol{Y},\boldsymbol{\Lambda})$  given by
\begin{eqnarray}\label{eq: function1}
\mathrm{S}(\boldsymbol{Y}) & \leq & \mathrm{G}(\boldsymbol{Y},\boldsymbol{\Lambda})=\sum_{i<j}\Lambda_{ij}\tilde{d}_{ij}^2+\sum_{i<j}\Lambda_{ij}d_{ij}^2(\boldsymbol{Y})\nonumber \\  & & -2\sum_{i<j}\Lambda_{ij}\tilde{d}_{ij}\frac{(\boldsymbol{y}_i-\boldsymbol{y}_j)^T (\boldsymbol{\Lambda}_i-\boldsymbol{\Lambda}_j)}{\parallel\boldsymbol{v_i}-\boldsymbol{v_j}\parallel}.
\end{eqnarray}
The function $\mathrm{G}(\boldsymbol{Y},\boldsymbol{\Lambda})$ can be put into matrix form as
\begin{equation}\label{eq: functionmatrix}
\mathrm{G}(\boldsymbol{Y},\boldsymbol{\Lambda})=\boldsymbol{B} + \text{Tr}(\boldsymbol{Y}^T\boldsymbol{Z}\boldsymbol{Y})-2\text{Tr}(\boldsymbol{Y}^T\boldsymbol{C}(\boldsymbol{\Lambda})\boldsymbol{\Lambda}).
\end{equation}
where $\text{Tr}(.)$ is the trace operator. The elements of matrices $\boldsymbol{B}$, $\boldsymbol{Z}$ and $\boldsymbol{C}$ are defined as
\begin{equation}\label{eq: B}
b_{ij} =
\begin{cases}
 \Lambda_{ij} \tilde{d}_{ij}^2, & \text{if $i\neq j$},\\
\\
0 ,& \text{if $i= j$},
 \end{cases}
\end{equation}
\begin{equation}\label{eq: Z}
z_{ij} =
\begin{cases}
\sum_{i=1,i\neq j}^{K} \Lambda_{ij}, & \text{if $i\neq j$},\\
\\
0 ,& \text{if $i= j$},
 \end{cases}
\end{equation}
and 
\begin{equation}\label{eq: C}
c_{ij} =
\begin{cases}
\sum_{i=1,i\neq j}^{K} \Lambda_{ij}\frac{\tilde{d}_{ij}}{d_{ij}(\boldsymbol{Y})}, & \text{if $i\neq j$},\\
\\
0, & \text{if $i= j$},
 \end{cases}
\end{equation}
respectively. Whereas $z_{ij}$ is the cumulative link qualities from all other nodes to node $j$, $c_{ij}$ is obtained by weighting $z_{ij}$ by the estimated distance normalized by the actual distance.  Obviously, function $\mathrm{G}(\boldsymbol{Y},\boldsymbol{\Lambda})$ is a quadratic function in $\boldsymbol{Y}$; therefore, analytically, its minimum is defined as
\begin{equation}\label{eq: stressmin}
\frac{\partial\mathrm{G}(\boldsymbol{Y},\boldsymbol{\Lambda})}{\boldsymbol{Y}}= 2\boldsymbol{Z}\boldsymbol{Y}-2 \boldsymbol{C}(\boldsymbol{\Lambda})\boldsymbol{\Lambda} = 0
\end{equation}
Finally, the estimated locations $\boldsymbol{\hat{Y}}$ are obtained from \eqref{eq: stressmin}, which minimizes $\mathrm{G}(\boldsymbol{Y},\boldsymbol{\Lambda})$ with respect to the actual locations $\mathbf{Y}$ as follows
\begin{equation}\label{eq: minimum}
\boldsymbol{\hat{Y}} =\boldsymbol{Z}^{-1}\boldsymbol{C}(\boldsymbol{\Lambda})\boldsymbol{\Lambda}.
\end{equation}
In order to estimate the actual location of nodes by using the anchors, centralized network localization techniques commonly apply a transformation method after the iterative majorization computation in \eqref{eq: minimum} \cite{Leeuw2009, nasir2015}. These approaches modify the actual location of anchors after solving a minimization problem, which eventually results in a low localization accuracy as the anchor positions are not fully utilized.  Moreover, they require a transformation based on the anchor locations as the majorization-approach-based location estimations, $\boldsymbol{\hat{Y}}$, are not globally estimated \cite{Leeuw2009}. However, a more practical solution is to include the anchor locations in the iterative majorization approach to get the global location of the nodes in the network. Therefore, we propose a modified iterative majorization formulation that improves the localization accuracy by directly including the anchor's location. The proposed approach does not require any additional transformation, in comparison with its counterpart. In order to approximate the missing range measurements, matrices $\boldsymbol{Y}$, $\boldsymbol{\Lambda}$, $\boldsymbol{C}(\boldsymbol{\Lambda})$ and $\boldsymbol{Z}$ in \eqref{eq: minimum} can partitioned as follows
\begin{equation}
\boldsymbol{Y}_{N_a} = [\boldsymbol{y}_1,...,\boldsymbol{y}_{N_a}] \in \mathbb{R}^{N_a\times \rho},
\end{equation}
\begin{equation}
\boldsymbol{Y}_{M} = [\boldsymbol{y}_1,...,\boldsymbol{y}_{M}] \in \mathbb{R}^{M\times \rho},
\end{equation}
\begin{equation}
\boldsymbol{\Lambda}_{N_a} = [\boldsymbol{\Lambda}_1,...,\boldsymbol{\Lambda}_{N_a}] \in \mathbb{R}^{N_a\times \rho},
\end{equation}
\begin{equation}
\boldsymbol{\Lambda}_{M} = [\boldsymbol{\Lambda}_1,...,\boldsymbol{\Lambda}_{M}] \in \mathbb{R}^{M\times \rho},
\end{equation}
\begin{equation}
\boldsymbol{C}(\boldsymbol{\Lambda}) = \begin{bmatrix}
\boldsymbol{C}_{11} &  \boldsymbol{C}_{12} \\ \boldsymbol{C}_{21} & \boldsymbol{C}_{22}
\end{bmatrix},
\end{equation}
and
\begin{equation}
\boldsymbol{Z} = \begin{bmatrix}
\boldsymbol{Z}_{11} &  \boldsymbol{Z}_{12} \\ \boldsymbol{Z}_{21} & \boldsymbol{Z}_{22}
\end{bmatrix},
\end{equation}
where the block matrices $\boldsymbol{C}_{11}$/$\boldsymbol{Z}_{11}$ is of length $N_a \times N_a$ and related to node-to-node elements, $\boldsymbol{C}_{12}$/$\boldsymbol{Z}_{12}$ is $M\times N_a$ long and corresponds to anchor-to-node elements, and the length of $\boldsymbol{C}_{22}$/$\boldsymbol{Z}_{22}$ is  $M\times M$ and is related to anchor-to-anchor elements. The missing elements for the block matrices\footnote{Element $c_{ij}$ is missed if there is no connectivity thus no signal reception from node $i$ to node $j$} $\boldsymbol{C}_{11}$ and $\boldsymbol{C}_{12}$  are  approximated by 	
\begin{equation}\label{eq: minmax}
\acute{c}_{ij} = \frac{\acute{R}_{min}+\acute{R}_{max}}{2},
\end{equation}
where $\acute{R}_{max}$ is the longest edge in the network and $\acute{R}_{min}$ is the shortest edge. Note that the approach in \eqref{eq: minmax} is different from the shortest path distance estimation. In \cite{Kashniyal2017, Shang, nasir2015}, the authors considered the shortest path estimation for the missing elements of the kernel matrix. Nevertheless, the shortest path distance estimation leads to a large error due to the accumulated error over the multi-hop path. The shortest path estimation also depends on the distribution of the sensor nodes in the network; if nodes are densely and uniformly deployed, then, the shortest path can be a good approximation. However, in the case of sparsely deployed nodes, it is not an optimal solution.  Therefore, in this paper, we propose a novel technique to approximate the missing elements for the block kernel matrix. By evaluating \eqref{eq: functionmatrix} in terms of the partitioned matrices, we obtain
\begin{align}\label{eq: partition}
\mathrm{G}(\boldsymbol{Y},\boldsymbol{\Lambda})&=\boldsymbol{B} + \sum_{k=1}^{K}\left(\boldsymbol{Y}^T_{N_{a(k)}}\boldsymbol{Z}_{11}\boldsymbol{Y}_{N_{a(k)}} \right. \nonumber\\ & \left. +2\boldsymbol{Y}^T_{N_{a(k)}}\boldsymbol{Z}_{12}\boldsymbol{Y}_{M_{(k)}}+\boldsymbol{Y}^T_{M_{(k)}}\boldsymbol{Z}_{22}\boldsymbol{Y}_{M_{(k)}}\right) \nonumber \\
 & -2 \sum_{k=1}^{K}\left(\boldsymbol{Y}^T_{N_{a(k)}}\boldsymbol{C}_{11}\boldsymbol{\Lambda}_{N_{a(k)}} \right.  \nonumber \\
& \left. +2\boldsymbol{Y}^T_{N_{a(k)}}\boldsymbol{C}_{12}\boldsymbol{\Lambda}_{M_{(k)}}+\boldsymbol{Y}^T_{M_{(k)}}\boldsymbol{C}_{22}\boldsymbol{\Lambda}_{M_{(k)}}\right).
\end{align}

Differentiating  \eqref{eq: partition} with respect $\boldsymbol{Y}_{N_{a(k)}}$ 
\begin{eqnarray}\label{eq: estimation}
\frac{\partial \mathrm{G}(\boldsymbol{Y},\boldsymbol{\Lambda})}{\boldsymbol{Y}_{N_{a(k)}}}&=& 2(\boldsymbol{Z}_{11}\boldsymbol{Y}_{N_{a(k)}}+\boldsymbol{Z}_{12}\boldsymbol{Y}_{M_{(k)}}\\
& & -\boldsymbol{C}_{11}\boldsymbol{\Lambda}_{N_{a(k)}}-\boldsymbol{C}_{12}\boldsymbol{\Lambda}_{M_{(k)}}).\nonumber
\end{eqnarray}
By setting \eqref{eq: estimation} equal to zero, the location estimation $\boldsymbol{\hat{Y}}_{N_{a(k)}}$ of the $N_a$ sensor nodes are given as
\begin{equation}
\boldsymbol{\hat{Y}}_{N_{a(k)}}=\boldsymbol{Z}_{11}^{-1}(\boldsymbol{C}_{11}\boldsymbol{\Lambda}_{N_{a(k)}}+\boldsymbol{C}_{12}\boldsymbol{Y}_{M_{(k)}}+\boldsymbol{Z}_{12}\boldsymbol{Y}_{M_{(k)}}),
\end{equation}
which can also be put into matrix form as
\begin{equation}\label{eq: solution}
\boldsymbol{\hat{Y}}_{N_a}=\boldsymbol{Z}_{11}^{-1}(\boldsymbol{C}_{11}\boldsymbol{\Lambda}_{N_a}+\boldsymbol{C}_{12}\boldsymbol{Y}_{M}+\boldsymbol{Z}_{12}\boldsymbol{Y}_{M})
\end{equation}
Notice that the solution for additional sensor nodes that are activated by the energy-harvesting source is straightforward from \eqref{eq: solution}, by adding $N_p$ number of nodes to $N_a$.
\subsection{Energy Cost of the Pairwise Measurements}
In network localization methods, the energy cost of transmitting the pairwise distance measurements to the surface node can be modeled as follows \cite{Rabbat2004}
\begin{equation}
E(K) = b(K)\times h(K) \times e(K),
\end{equation}
where $b(K)$ represents the number of bits per packet transmitted, $h(K)$ is the node degree, and $e(K)$ is the required energy to transmit a single bit. Then the number of packets transmitted to the surface node is $b_s(K) = \rho m K$, where $\rho$ is the ratio of the packet size to measurement size and $m$ is the average number of neighbors. In two-dimensional space, the average number of hops to the surface node is $h_s(K) = O(\sqrt{K})$. Since we have considered a multihop network setup, the total energy cost of all the nodes to transmit the pairwise distances to the surface node is given as
\begin{equation}
E_s(K) = \rho m K^{\frac{3}{2}}\times e(K).
\end{equation}
\begin{figure}
\begin{center}
\includegraphics[width=0.99\columnwidth]{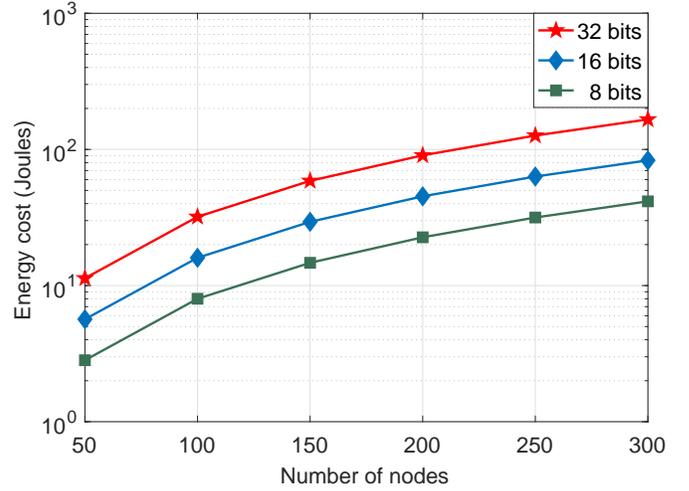}
\caption{Energy cost Vs. Number of nodes.}
\label{fig:energycost}
\end{center}
\end{figure}
Fig.~\ref{fig:energycost} shows the total energy cost of the network to communicate the pairwise distances to the surface node. It is clear that increasing both the number of nodes in the network and packet size, increases the energy cost to communicate the pairwise distances to the surface node in the network.

\section{Performance Comparison to CRLB Benchmark}\label{sec:analysis}
Graph realization and network localization are two closely related phenomena because true realization of a network can generally be obtained using is globally rigid graph\cite{Aspnes2006}. Nonetheless, the realization can still fail, even for globally rigid graphs, if the pairwise distances are corrupted by noise, which leads to the problem of flipping ambiguity \cite{Moore2004} and prevents the unique localization of a node. In order to avoid the flipping ambiguity, the anchor nodes in the network
should be well-separated and non-collinear\footnote{In this paper, we have considered a graph realization based network localization approach which is also subject to the flipping ambiguity and methods from \cite{Moore2004,Kashniyal20172,Kannan2007} can be used to overcome this problem.}. The localization of energy-harvesting UOWSN is similar to the parameter estimation problem. Therefore, to evaluate any parameter-estimation problem, the minimum unbiased variance estimation is taken as the evaluation criteria. Thus, the Cramer-Rao lower bound (CRLB) is commonly used as an unbiased parameter estimator to evaluate the performance of the parameter estimation \cite{Larssoncrlb2004}. 
\begin{figure*}
    \centering
    \begin{subfigure}[b]{0.45\textwidth}
\includegraphics[width=1\columnwidth]{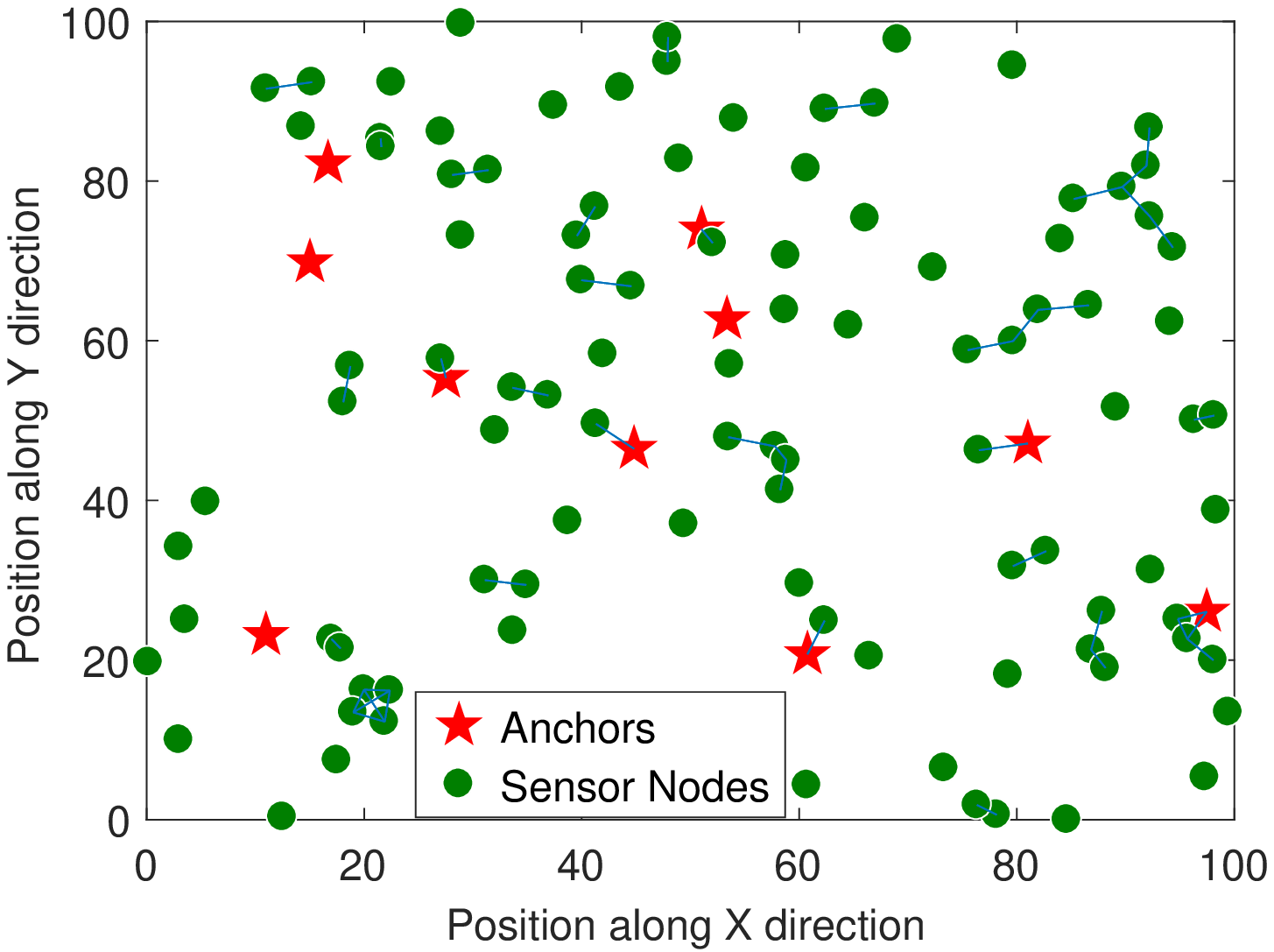}  
\caption{$N_a = 10,$  $N_p = 80$ and $M = 10$.\label{fig:20nodes}}  
    \end{subfigure}
    \begin{subfigure}[b]{0.45\textwidth}
\includegraphics[width=1\columnwidth]{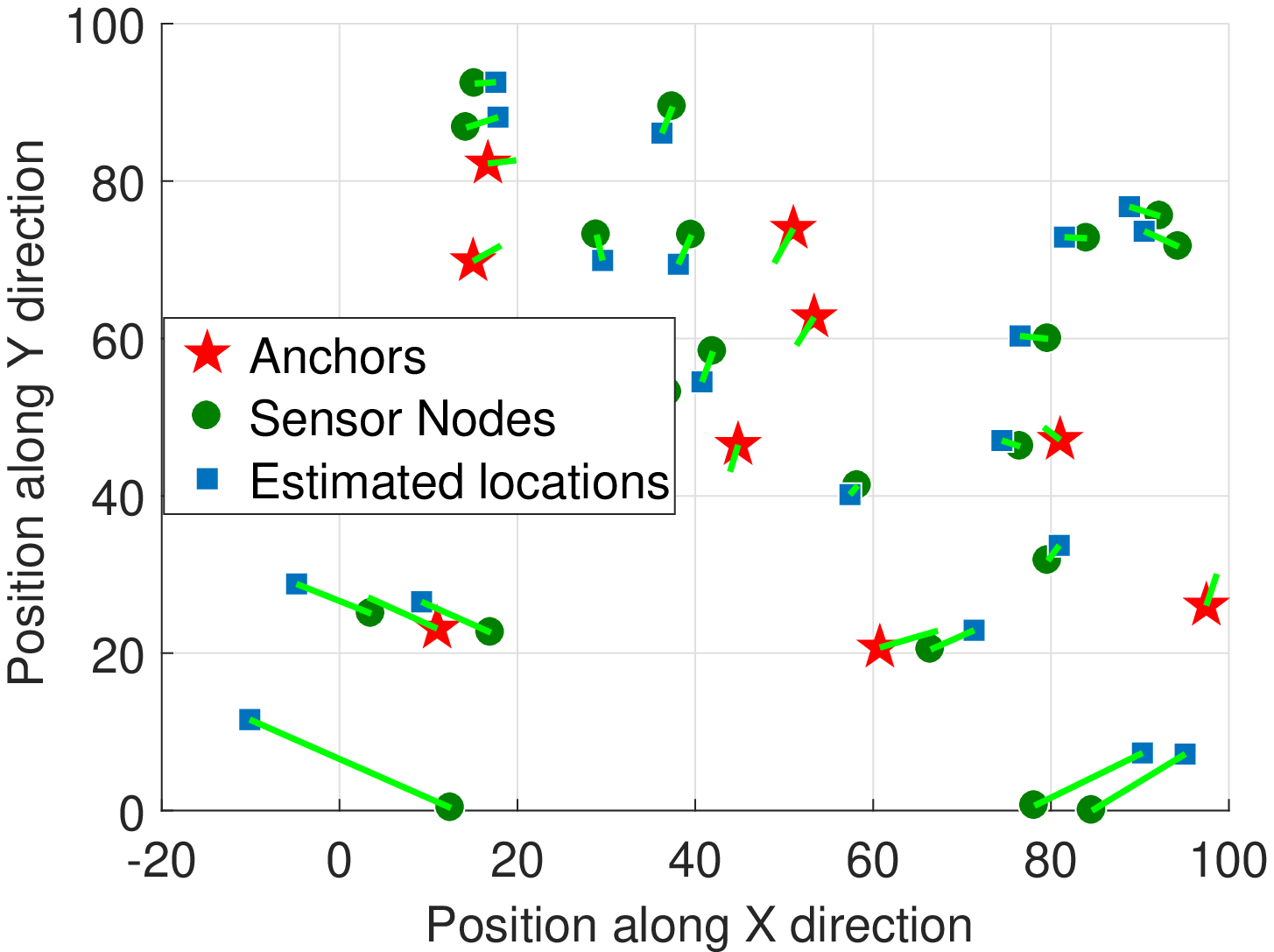}  
\caption{$N_a = 20,~ N_p = 70~\text{and}~M = 10$ .\label{fig:30nodesresult}}  
    \end{subfigure}
    
    \begin{subfigure}[b]{0.45\textwidth}
\includegraphics[width=1\columnwidth]{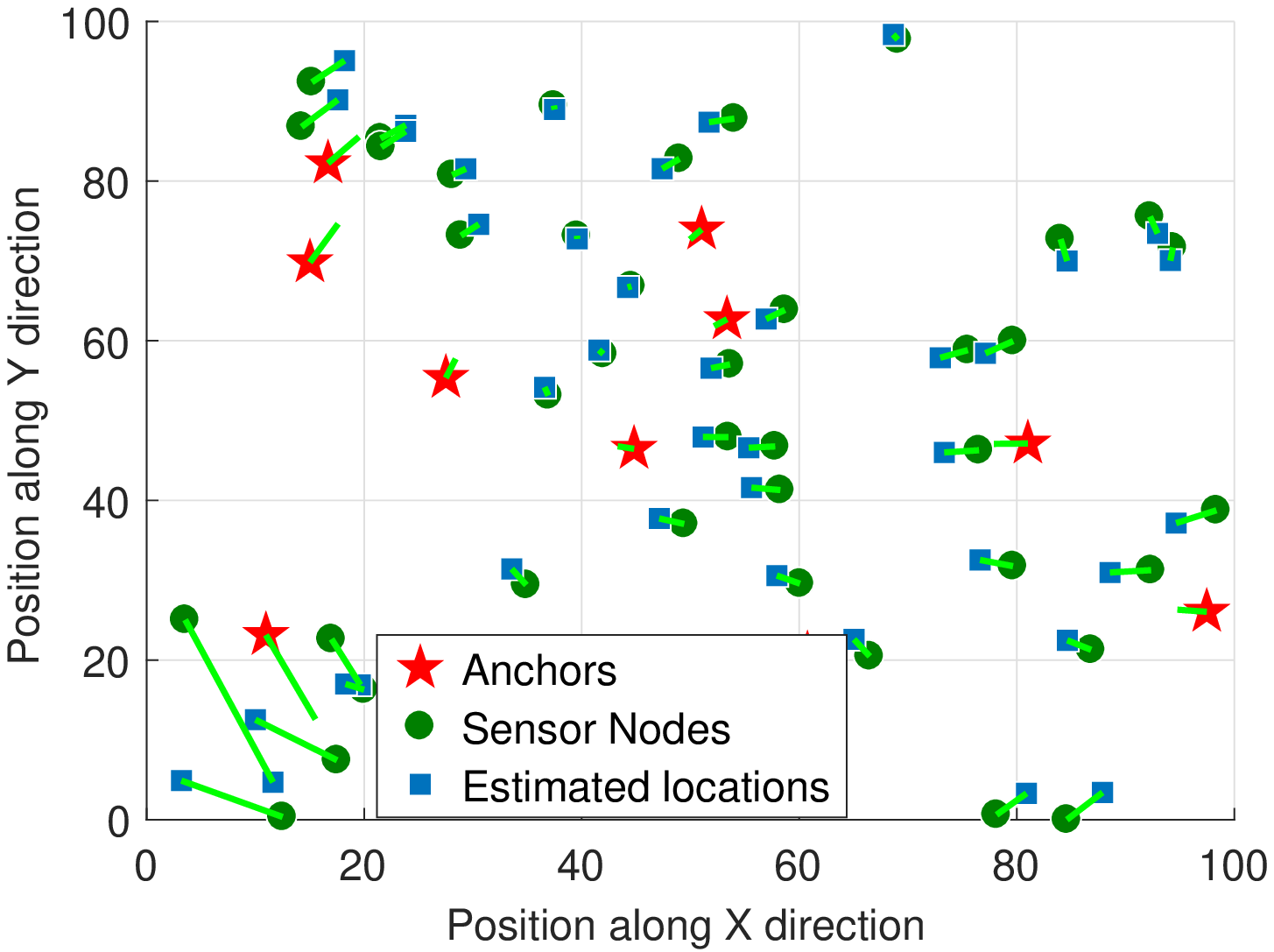}  
\caption{$N_a = 40,~ N_p = 50~\text{and}~M = 10$ .\label{fig:50nodesresult}}  
    \end{subfigure}
    \begin{subfigure}[b]{0.45\textwidth}
\includegraphics[width=1\columnwidth]{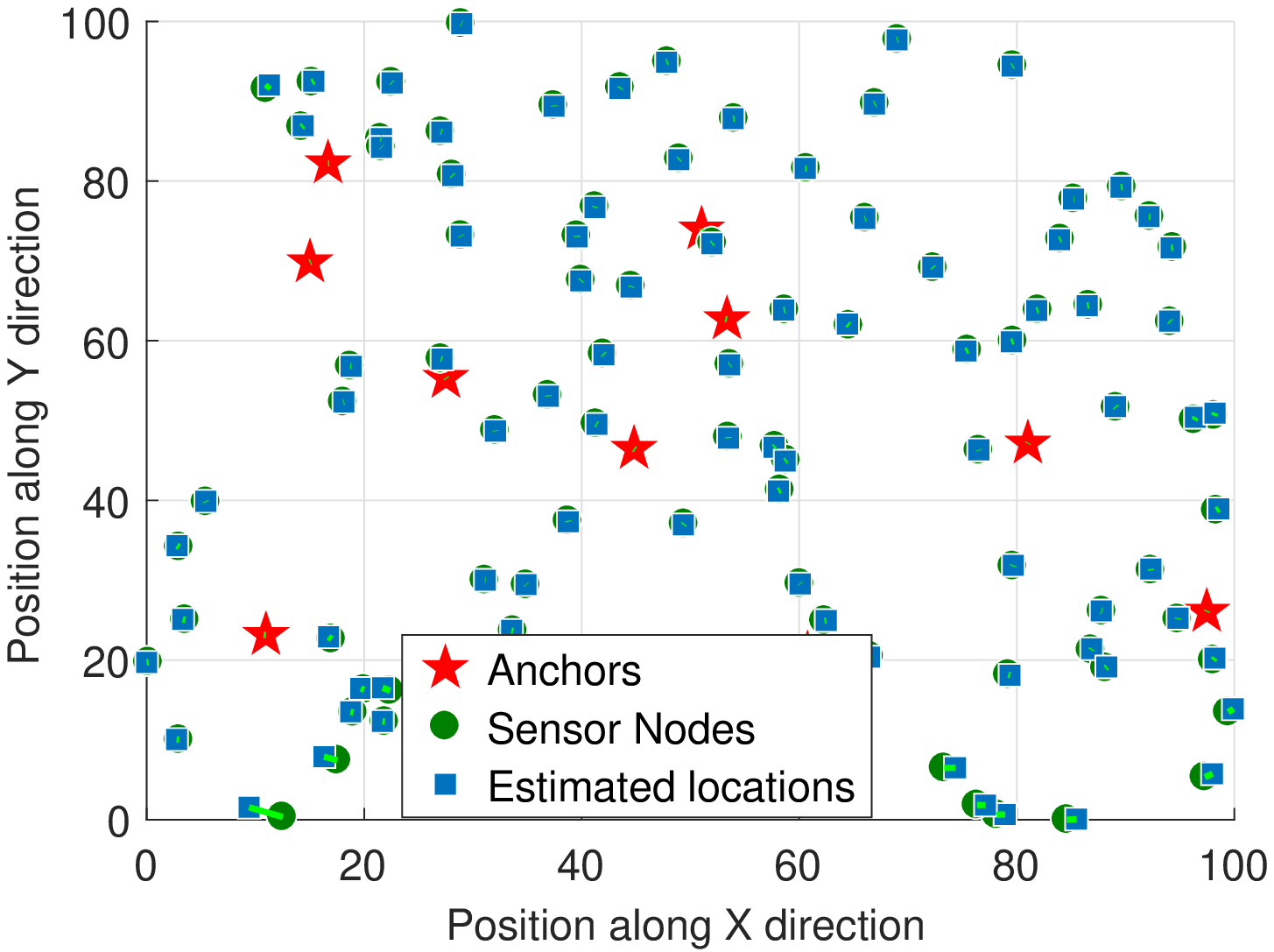}  
\caption{$N_a = 90,~ N_p = 0~\text{and}~M = 10$.\label{fig:90nodesresult}}  
    \end{subfigure}    
\caption{Localization error vs. number of active nodes (a) Network is not connected and therefore, cannot be localized, (b) RMSPE = 5.45 m, (c)  RMSPE = 1.22 m, (d) RMSPE = 0.22 m.}
\label{fig:rmsevsnodes_scenarios}
\vspace{4pt}
\hrule
\end{figure*}
Based on the noisy RSS measurement model in \eqref{eq: pr2}, the probability density function (PDF) of the received power is given as
\begin{equation}\label{eq: pdf}
f(\tilde{P}_{r_{ij}}|\boldsymbol{y}_i, \boldsymbol{y}_j)= \frac{1}{\sigma_{ij}\sqrt{2\pi}}\exp^{\left(-\frac{({\tilde{P}_{r_{ij}}-P_{r_{ij}}})^2}{2\sigma_{ij}^2}\right)}.
\end{equation}  
Accordingly, the corresponding log-likelihood ratio for the given PDF in \eqref{eq: pdf} can be written as
\begin{eqnarray}\label{eq: pdf2}
\vartheta_{ij}\text{[dB]} & = & -\log({\sigma_{ij}}\sqrt{2{\pi}})+\log\bigg(\exp^{-\frac{(\tilde{P}_{r_{ij}}-P_{r_j})^2}{2\sigma_{ij}^2}}\bigg).
\end{eqnarray}
The joint log-likelihood ratio for all the nodes in the network is
\begin{equation}\label{eq: jpdf}
\boldsymbol{\Theta} = \sum_{i=1}^{N_a}\sum_{j=i+1}^{K}\log\left(f(\tilde{P}_{r_{ij}}|\boldsymbol{y}_i, \boldsymbol{y}_j)\right),
\end{equation}
where $K = M+N_a$. The CRLB depends on the Fisher information matrix (FIM) that is derived from  the second-order derivative of the joint log likelihood function defined in \eqref{eq: jpdf}. The sub-matrices of FIM $\boldsymbol{{F}}$ are given as 
\begin{equation}\label{eq: FIM2}
\boldsymbol{{F}}= \begin{bmatrix}
\boldsymbol{{F}}_{2i-1,2i-1} & \boldsymbol{\mathcal{F}}_{2i-1,2i}\\ \boldsymbol{{F}}_{2i,2i-1} & \boldsymbol{{F}}_{2i,2i} \end{bmatrix}, \:\: i=1,2,3......K.
\end{equation}
The elements of the sub-matrices are derived in appendix A. Moreover, the diagonal elements of the FIM are given as
\begin{equation}\label{eq: FIM3}
\begin{split}
&\boldsymbol{{F}}_{{xx}_{2i-1,2j-1}}  =  \boldsymbol{{F}}_{{xx}_{2j-1,2i-1}} = \\ & ~~ 
\begin{cases}
 \sum_{j\in{H(i)}} \frac{3 k_{ij} \mu}{\sigma_{ij}^2 d_{ij}^5}\bigg(1-\frac{5(x_i-x_j)^2}{d_{ij}^2}\bigg)  & \text{ $i=j$},\\
 -\frac{3 k_{ij} \mu}{\sigma_{ij}^2 d_{ij}^5}\bigg(1-\frac{5(x_i-x_j)^2}{d_{ij}^2}\bigg) & \text{$j\in{H(i)}$ and $j\neq i$}\\
 0 & \text{$j\not\in{H(i)}$}
 \end{cases},
\end{split}
\end{equation}
\begin{equation}\label{eq: FIM5}
\begin{split}
& \boldsymbol{{F}}_{{yy}_{2j,2i}} = \\ & ~~ 
\begin{cases}
 \sum_{j\in{H(i)}}\frac{3 k_{ij} \mu}{\sigma_{ij}^2 d_{ij}^5}\bigg(1-\frac{5(y_i-y_j)^2}{d_{ij}^2}\bigg)  & \text{ $i=j$},\\
 -\frac{3 k_{ij} \mu}{\sigma_{ij}^2 d_{ij}^5}\bigg(1-\frac{5(y_i-y_j)^2}{d_{ij}^2}\bigg) & \text{$j\in{H(i)}$ and $j\neq i$}\\
 0 & \text{$j\not\in{H(i)}$}
 \end{cases},
 \end{split}
\end{equation}
where $\mu = \exp^{-e(\lambda)}$ and $k_{ij}=P_{t_i}\eta_{i}\eta_j\frac{A_j\cos\theta}{2\pi(1-\cos\theta_0)}$. On the other hand, the non-diagonal elements of the FIM are given as
\begin{equation}\label{eq: FIM6}
\begin{split}
 & \boldsymbol{{F}}_{{xy}_{2i-1,2j}} = \boldsymbol{{F}}_{{xy}_{2j,2i-1}} =\boldsymbol{{F}}_{{xy}_{2i,2j-1}} =   \boldsymbol{{F}}_{{xy}_{2j-1,2i}}= \\ & 
\begin{cases}
 \sum_{j\in{H(i)}}-\frac{15 k_{ij} \exp^{-e(\lambda)}(x_i-x_j)(y_i-y_j)}{\sigma_{ij}^2 d_{ij}^7} & \text{ $i=j$},\\
\frac{15 k_{ij} \exp^{-e(\lambda)}(x_i-x_j)(y_i-y_j)}{\sigma_{ij}^2 d_{ij}^7} & \text{$j\in{H(i)}$, $j\neq i$}\\
 0 & \text{$j\not\in{H(i)}$}
 \end{cases}.
 \end{split}
\end{equation}
Then, the CRLB is defined as
\begin{equation}
\text{CRLB} = \text{Tr}(\boldsymbol{{F}}^{-1}),
\end{equation}
which accounts for the different properties of underwater optical channel characteristics and the noise error variance of the receiver. The RMSPE of the proposed localization method should satisfy the following condition
\begin{equation}
\text{CRLB} \leq \sqrt{\frac{\sum_{i=1}^{N_a} (\hat{x}_i-x_i)^2+(\hat{y}_i-y_i)^2}{N_a}}.
\end{equation}

\begin{figure}[t]
\begin{center}
\includegraphics[width=0.99\columnwidth]{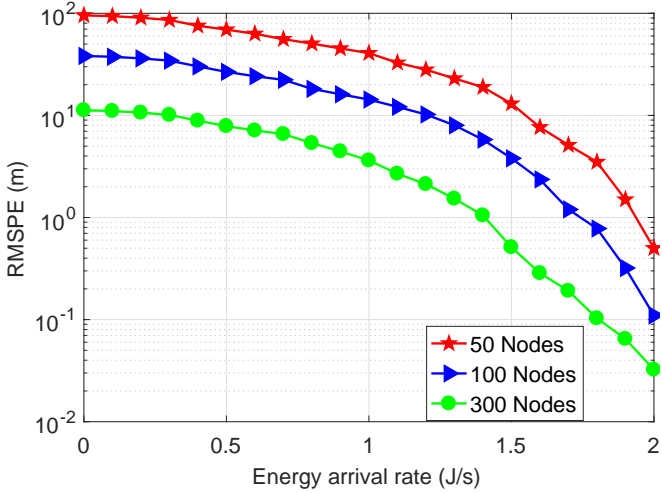}
\caption{RMSPE Vs. Energy arrival rate with fixed number of active nodes.}
\label{fig:energyarrivall}
\end{center}
\end{figure}
\section{Simulation Results}\label{sec:results}

A number of Monte Carlo simulations are conducted in MATLAB to analyze the performance of the proposed approach. Initially, we consider 100 nodes randomly distributed in a $100 \times 100 \:m^2$ area, where active nodes $N_a$ are 10, passive nodes $N_p$ are 80, and anchor nodes $M$ are 10, as shown in Fig.~4. As a reminder, the degree of connectivity increases with the network density, which eventually improves the network localization accuracy. Therefore, the performance of the proposed localization algorithm primarily depends on the node density regardless of the network area. In practice, energy arrivals to the harvesters have an intermittent nature, and the passive nodes can harvest energy from the ambient energy sources and get connected to the network. 

We consider a generic multi-source energy-harvesting model that collects energy from both microbial fuel cells \cite{mfc2015} and acoustic harvesters \cite{Li2016}. Experimental results show that I-V curve of MFCs can provide a maximum power point (MPP) up to 1.5 Watts per quarter meter quare (\cite{RAHIMNEJAD2015745} and references therein). On the other hand, it is possible to equip the surface of the sensor with multiple piezoelectric cantilever beams and to combine the generated acoustic energy to achieve a sufficient sorce of power \cite{liu2008acoustic}. Unfortunately, to the best of our knowledge, statistics of energy arrivals in an underwater environment have not been investigated yet  for these types of energy harvester. Therefore, we assume that the combined energy arrival rate to each node is uniformly distributed between $0.5$ and $2$ Joules per second. 

For the duty cycle optimization, we use the following parameters, $\mathcal{T}=1000$, $T=1$ seconds, $\beta=0.1$, $B_0=1$ Joules, and $B_{\max}=10$ Joules. Contrary to the long-term observation history-based average energy harvesting rates that are randomly chosen between $0$ and $2$ Joules per second, the power consumption $P_c^i$ is calculated to ensure that the transmission capability of each sensor node is kept constant at 20 m, in clear ocean water, with the absorption coefficient of $a(\lambda)=0.114$ and scattering coefficient of $s(\lambda)=0.037$.  Undoubtedly, the ranging error has a negative affect on the accuracy of every localization method. Therefore, we assume that the ranging errors are Gaussian-distributed, with zero mean and variance $\sigma^2$, with a value of $\sigma^2$ set to $0.02$ m for the rest of the simulations, except those in which the impact of the variable $\sigma^2$ is investigated (Fig. 11).

\begin{figure}[t]
\begin{center} 
\includegraphics[width=0.99\columnwidth]{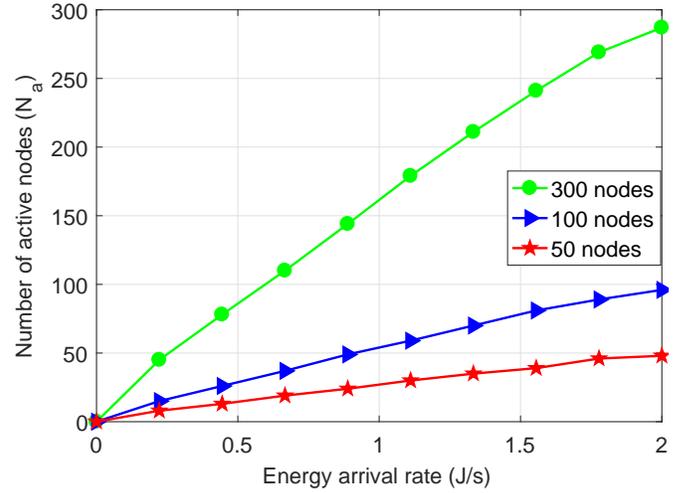}
\caption{RMSPE Vs. Energy arrival rate with fixed transmission range.}
\label{fig:noofactivnodesvsenergyarrival}
\end{center}   
\end{figure} 
 
Fig.~\ref{fig:20nodes} clearly shows that, due to the low-energy arrival rates, most of the nodes are in passive mode and thus disconnected from the network. As a result of the limited network connectivity, the proposed network localization approach cannot determine the node positions, as it requires more active nodes to estimate the entire network. Therefore, the harvested energy improves the number of active nodes, and once the network is connected, it is possible to localize all the nodes in the network  where 10-80 of the passive nodes harvest energy from ambient energy sources and get connected to the network (Fig.~\ref{fig:30nodesresult}~-~Fig.~\ref{fig:90nodesresult}).  Fig.~\ref{fig:30nodesresult}~-~Fig.~\ref{fig:90nodesresult} also show that, when the active nodes are increased from 10 to 90, i.e., all the nodes in the network are active, the localization performance is significantly improved, and RMSPE is reduced by 30\%.

\begin{figure}
\begin{center} 
\includegraphics[width=0.99\columnwidth]{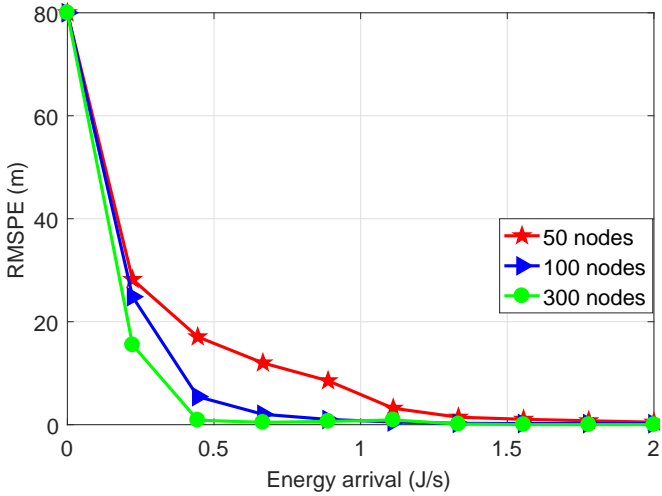}
\caption{Number of active nodes Vs. Energy arrival rate.}
\label{fig:rmsevsenergyarrivalrate}
\end{center}   
\end{figure}
\begin{figure}
\begin{center} 
\includegraphics[width=0.99\columnwidth]{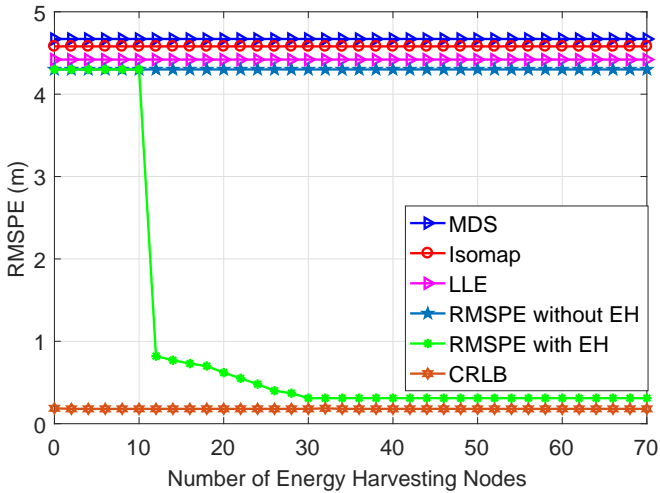}
\caption{RMSPE Vs. $N_a$ }
\label{fig:rmspevsns}
\end{center}   
\end{figure}

In this study, we have also investigated the impact of the energy arrival rate on the localization performance, as shown in Figs. \ref{fig:energyarrivall} to  \ref{fig:rmsevsenergyarrivalrate}. Fig. \ref{fig:energyarrivall} presents the simulations for three different scenarios with 50, 100, and 300 nodes, each randomly distributed in a square region of 100 m $\times$ 100 m. The maximum energy arriving into the network is up to 2 joules/sec, as per the experimental results of microbial fuel cells and acoustic harvester.  A number of Monte Carlo simulations are performed for each scenario. Fig. \ref{fig:energyarrivall} shows that the transmission range of active nodes increases as the energy arrival rate increases, leading to a reduction in the RMSPE. Fig. \ref{fig:energyarrivall} also shows that the denser networks can provide a better RMSPE,  even for low energy arrival rates. Additionally, based on duty-cycle optimization, energy harvesting increases the number of active nodes in the network and improves the connectivity of the network and the RMSPE, as shown in Figs. 5 to 7.

\begin{figure}
\begin{center} 
\includegraphics[width=0.99\columnwidth]{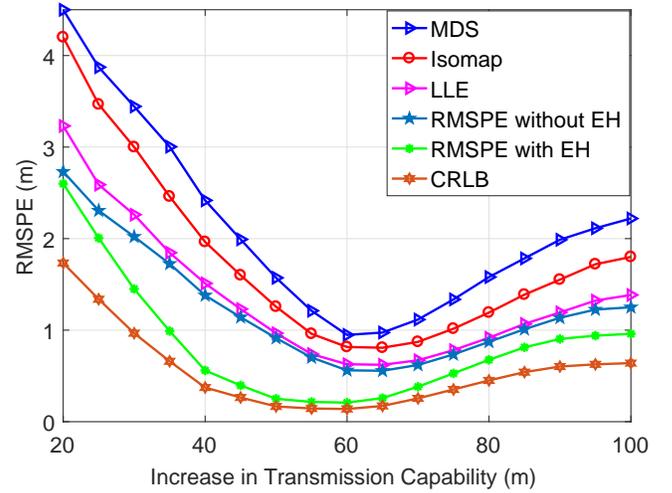}
\caption{RMSPE Vs. Transmission Capability }
\label{fig:rmspevstx}
\end{center}   
\end{figure}
\begin{figure}
\begin{center} 
\includegraphics[width=0.99\columnwidth]{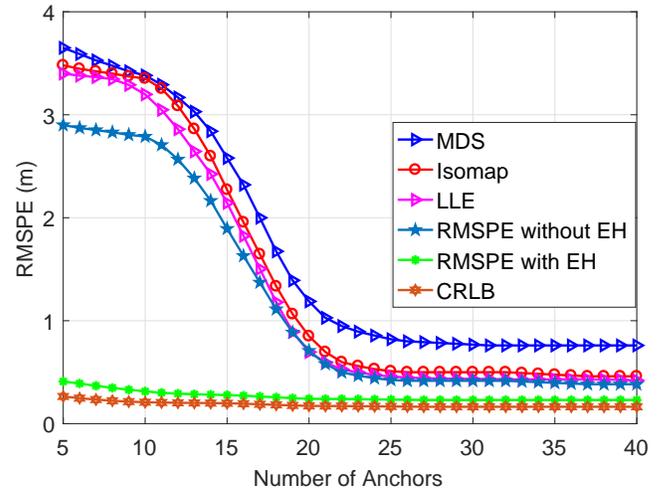}
\caption{RMSPE Vs. Number of Anchors}
\label{fig:rmspevsan}
\end{center}   
\end{figure} 

Figs.~\ref {fig:rmspevsns} to \ref {fig:rmspevsan}, show that, in comparison with well-known network localization techniques such as Isomap \cite{Kashniyal2017}, multidimensional scaling (MDS) \cite{Shang, nasir2015} and local linear embedding (LLE) \cite{Roweis2323}, the proposed network localization technique  provide better results due to energy harvesting, block kernel matrices, and a better shortest path estimation. Fig.~\ref {fig:rmspevsns} shows the relationship between energy harvesting and RMSPE. Energy harvesting increases the number of active nodes in the network, thus reduces the multi-hop errors in the network. The RMSPE is also compared to well-known network localization techniques such as Isomap \cite{Kashniyal2017}, multidimensional scaling (MDS) \cite{Shang, nasir2015} and local linear embedding (LLE) \cite{Roweis2323}. Fig.~\ref {fig:rmspevsns} shows that, when the network is connected, the proposed method achieves the CRLB. In Fig~.\ref{fig:rmspevstx}, the RMSPE performance of the proposed EH-UOWSNs localization is compared with Isomap, MDS, and LLE in terms of transmission range capability of the sensor nodes. The RMSPE improves with the increase in the transmission range, up to a level of approximately 60 m, above which it starts to increase due to the flipping ambiguity.

\begin{figure}
\begin{center} 
\includegraphics[width=0.99\columnwidth]{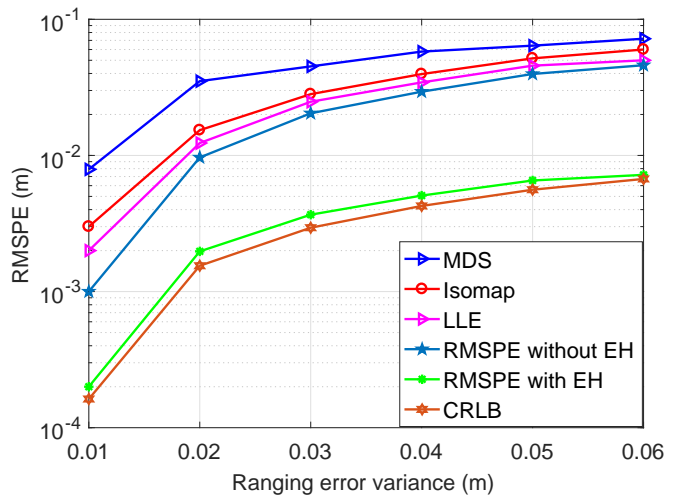}
\caption{RMSPE Vs. Ranging error variance}
\label{fig:rmspevsnoise}
\end{center}   
\end{figure}

Fig.~\ref{fig:rmspevsan} shows the impact of increasing the number of anchors in a network containing 90 sensor nodes deployed in a $100 \times 100~ m^2$  area, and with a range of communication set at 20 m.  Fig.~\ref{fig:rmspevsan} clearly shows that, due to the energy-harvesting capabilities, block kernel matrices and better shortest path estimation of the proposed technique outperforms Isomap, MDS, and LLE. Undoubtedly, the ranging error and the estimation of missing pairwise distances have a negative effect on the accuracy of every localization system. As hypothesized that the ranging errors are Gaussian-distributed, with zero mean and variance $\sigma^2$. Here, we examine the performance of the proposed technique with respect to the variable $\sigma^2$, i.e.,  $\sigma^2 = 0.01-0.06$ m. To examine the impact of the ranging error, we consider 100 optical sensor nodes and 10 anchor nodes, randomly deployed in a $100 \times 100 \:m^2$ area, with the transmission range of 40 m. Note that the results are averaged over 100 different network setups. Fig.~\ref{fig:rmspevsnoise} shows that the proposed localization system is more robust to the ranging error than Isomap, MDS, and LLE, mainly because of its better approximation of missing pairwise distances.

\begin{figure}
\begin{center} 
\includegraphics[width=1\columnwidth]{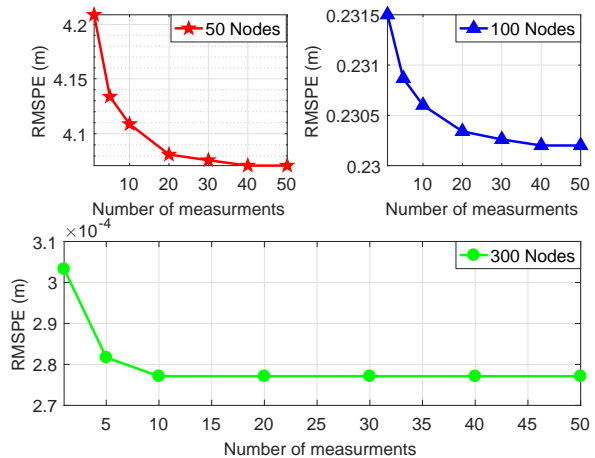}
\caption{RMSPE Vs. Number of measurements}
\label{fig:rmspevsmeasurements}
\end{center}   
\end{figure}

The impact of the temporal nature to get multiple RSS measurements each with different noise realizations is investigated in Fig.~\ref{fig:rmspevsmeasurements}. Three different network setups are simulated  with 50, 100, and 300 nodes randomly distributed in a square region of 100 m $\times$ 100 m. The transmission range of the nodes for each setup is kept constant, at 40 m. Fig.~\ref{fig:rmspevsmeasurements} shows that increasing the number of measurements for each pairwise distance estimation improves the accuracy, up to certain number of measurements, after which the RMSPE gets saturated.

\section{Conclusions}\label{sec:conc}
In this paper, an energy-harvesting-based localization technique is developed for underwater optical sensor networks, using RSS measurements.  In an aquatic environment, its difficult to replace or recharge the battery of a sensor node. Therefore, designing an efficient and reliable energy harvester for the continuous operation of UOWSN is required. We develop a mathematical model that can harvest energy from multiple sources and distribute it to the sensor nodes.  The RSS measurements for underwater optical communications are inaccurate and lead large localization errors. The proposed technique takes into account the energy from the energy harvesting sources, thus making it more robust than other network localization techniques. The proposed method also reduces errors in estimating the shortest paths, in block kernel matrices, by introducing a novel matrix completion technique. Additionally, the CRLB is derived for the proposed EH-UOWSN localization technique. Simulations show that the proposed technique for underwater optical sensor networks localization is a good strategy to get robust and accurate results.

\appendices
\section{Derivation of The FIM Elements}
Since the CRLB depends on the FIM, the elements of FIM are derived from the second-order partial differentiation of the likelihood function as follows
\begin{equation}\label{eq: fima1}
\boldsymbol{{F}}_{xx_{ij}} = -E\bigg(\frac{\partial^2 \ln f(\tilde{P}_{r_{ij}}|\boldsymbol{y}_i, \boldsymbol{y}_j)}{\partial x_i^2}\bigg),
\end{equation}
\begin{equation}\label{eq: fima2}
\boldsymbol{{F}}_{yy_{ij}} = -E\bigg(\frac{\partial^2 \ln f(\tilde{P}_{r_{ij}}|\boldsymbol{y}_i, \boldsymbol{y}_j)}{ \partial y_i^2}\bigg), 
\end{equation}
\begin{equation}\label{eq: fima3}
\text{and}~ \boldsymbol{{F}}_{xy_{ij}} = -E\bigg(\frac{\partial^2 \ln f(\tilde{P}_{r_{ij}}|\boldsymbol{y}_i, \boldsymbol{y}_j)}{\partial x_i \partial y_i}\bigg).
\end{equation}
The second-order partial differentiation of the likelihood function, with respect to the estimation parameters $x$ and $y$, are expressed as follows
\begin{align}
\nonumber  \frac{\partial}{\partial x_i} \ln f(\tilde{P}_{r_{ij}} &|\boldsymbol{y}_i, \boldsymbol{y}_j) = \frac{\partial}{\partial x_i} (-\log(\sigma_{ij}\sqrt{2\pi}))\\
&-\frac{\partial}{\partial x_i}\bigg(\frac{1}{\sigma_{ij}^2}(\tilde{P}_{r_{ij}}-\frac{k_{ij}\exp^{-e(\lambda)d_{ij}}}{d_{ij}^2})\bigg)
\end{align}
\begin{equation}
\frac{\partial}{\partial x_i} \ln f(\tilde{P}_{r_{ij}}|\boldsymbol{y}_i, \boldsymbol{y}_j) = \frac{-3 k_{ij} \exp^{-e(\lambda)}(x_i-x_j)}{\sigma_{ij}^2 d_{ij}^5}
\end{equation}
\begin{equation}\label{eq: fima4}
\frac{\partial^2}{\partial x_i^2} \ln f(\tilde{P}_{r_{ij}}|\boldsymbol{y}_i, \boldsymbol{y}_j) = \frac{3 k_{ij} \exp^{-e(\lambda)}}{\sigma_{ij}^2 d_{ij}^5}\bigg(\frac{5(x_i-x_j)^2}{d_{ij}^2}-1\bigg),
\end{equation}
\begin{equation}\label{eq: fima5}
\frac{\partial^2}{\partial y_i^2} \ln f(\tilde{P}_{r_{ij}}|\boldsymbol{y}_i, \boldsymbol{y}_j) = \frac{3 k_{ij} \exp^{-e(\lambda)}}{\sigma_{ij}^2 d_{ij}^5}\bigg(\frac{5(y_i-y_j)^2}{d_{ij}^2}-1\bigg),
\end{equation}
\begin{equation}\label{eq: fima6}
\frac{\partial^2}{\partial x_i \partial y_i} \ln f(\tilde{P}_{r_{ij}}|\boldsymbol{y}_i, \boldsymbol{y}_j) = \frac{15 k_{ij} \exp^{-e(\lambda)}(x_i-x_j)(y_i-y_j)}{\sigma_{ij}^2 d_{ij}^7}.
\end{equation}
Substituting the values from (45)-(47) into (40)-(42) leads to the elements of the FIM which are given in (34).

\bibliographystyle{../bib/IEEEtran}
\bibliography{../bib/IEEEabrv,../bib/nasir_ref}
\vspace{-1.2cm}
\begin{IEEEbiography}[{\includegraphics[width=1in,height=1.25in]{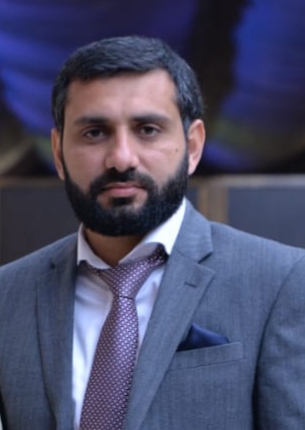}}]{Nasir Saeed}(S'14-M'16-SM'19) received his Bachelors of Telecommunication degree from University of Engineering and Technology, Peshawar, Pakistan, in 2009 and received Masters degree in satellite navigation from Polito di Torino, Italy, in 2012. He received his Ph.D. degree in electronics and communication engineering from Hanyang University, Seoul, South Korea in 2015. He was an assistant professor at the Department of Electrical Engineering, Gandhara Institute of Science and IT, Peshawar, Pakistan from August 2015 to September 2016. Dr. Saeed worked as an assistant professor at IQRA National University, Peshawar, Pakistan from October 2017 to July 2017. He is currently a postdoctoral research fellow at Communication Theory Lab, King Abdullah University of Science and Technology (KAUST).   His current areas of interest include cognitive radio networks, underwater optical wireless communications, dimensionality reduction, and localization.
\end{IEEEbiography}

\begin{IEEEbiography}[{\includegraphics[width=1in,height=1.25in]{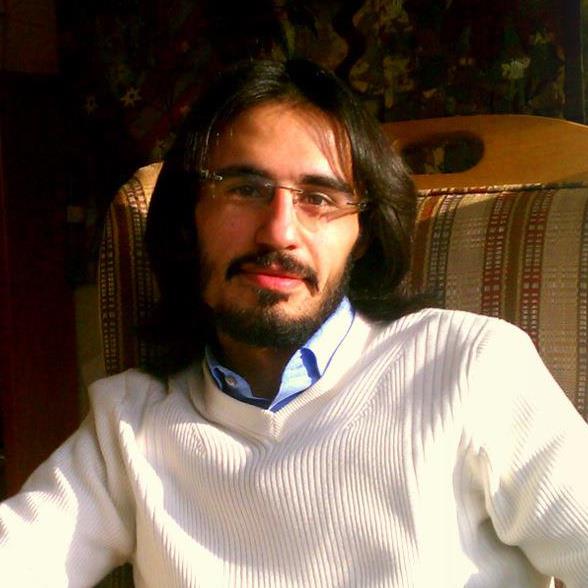}}]{Abdulkadir Celik}
(S'14-M'16) received the B.S. degree in electrical-electronics engineering from Selcuk University in 2009, the M.S. degree in electrical engineering in 2013, the M.S. degree in computer engineering in 2015, and the Ph.D. degree in co-majors of electrical engineering and computer engineering in 2016, all from Iowa State University, Ames, IA. He is currently a postdoctoral research fellow at Communication Theory Laboratory of King Abdullah University of Science and Technology (KAUST). His current research interests include but not limited to 5G networks and beyond, wireless data centers, UAV assisted cellular and IoT networks, and underwater optical wireless communications, networking, and localization. 
\end{IEEEbiography}

\begin{IEEEbiography}[{\includegraphics[width=1in,height=1.25in]{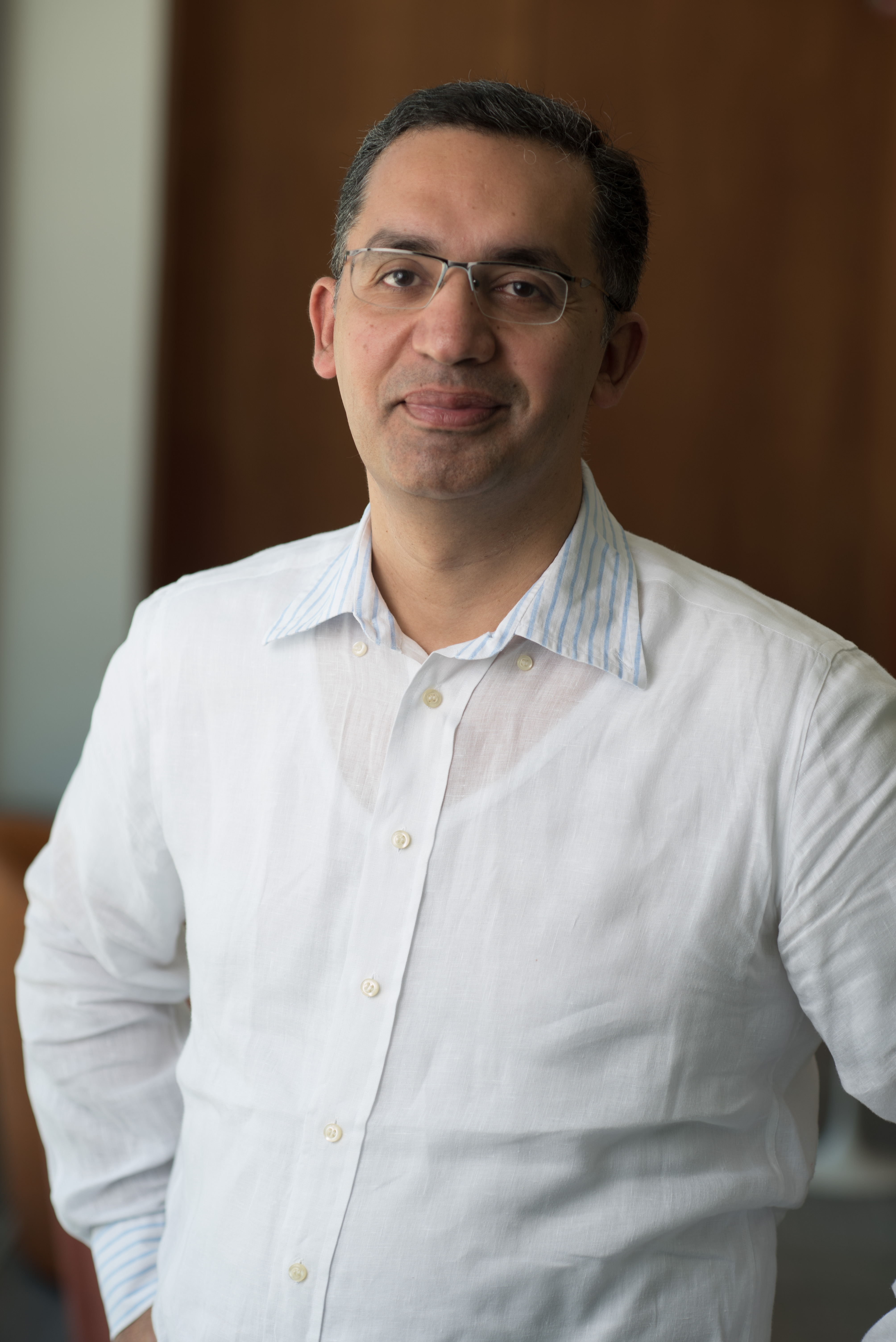}}]{Tareq Y. Al-Naffouri }
(M'10-SM'18) Tareq  Al-Naffouri  received  the  B.S.  degrees  in  mathematics  and  electrical  engineering  (with  first  honors)  from  King  Fahd  University  of  Petroleum  and  Minerals,  Dhahran,  Saudi  Arabia,  the  M.S.  degree  in  electrical  engineering  from  the  Georgia  Institute  of  Technology,  Atlanta,  in  1998,  and  the  Ph.D.  degree  in  electrical  engineering  from  Stanford  University,  Stanford,  CA,  in  2004.  

He  was  a  visiting  scholar  at  California  Institute  of  Technology,  Pasadena,  CA  in  2005  and  summer  2006.  He  was  a  Fulbright scholar  at  the  University  of  Southern  California  in  2008.  He  has  held  internship  positions  at  NEC  Research  Labs,  Tokyo,  Japan,  in  1998,  Adaptive  Systems  Lab,  University  of  California  at  Los  Angeles  in  1999,  National  Semiconductor,  Santa  Clara,  CA,  in  2001  and  2002,  and  Beceem  Communications  Santa  Clara,  CA,  in  2004.  He  is  currently  an  Associate Professor  at  the  Electrical  Engineering  Department,  King  Abdullah  University  of  Science  and  Technology  (KAUST).  His  research  interests  lie  in  the  areas  of  sparse, adaptive,  and  statistical  signal  processing  and  their  applications,  localization,  machine  learning,  and  network  information  theory.    He  has  over  240  publications  in  journal  and  conference  proceedings,  9  standard  contributions,  14  issued  patents,  and  8  pending. 

Dr.  Al-Naffouri  is  the  recipient  of  the  IEEE  Education  Society  Chapter  Achievement  Award  in  2008  and  Al-Marai  Award  for  innovative  research  in  communication  in  2009.  Dr.  Al-Naffouri  has  also  been  serving  as  an  Associate  Editor  of  Transactions  on  Signal  Processing  since  August  2013. 
\end{IEEEbiography}

\begin{IEEEbiography}[{\includegraphics[width=1in,height=1.25in]{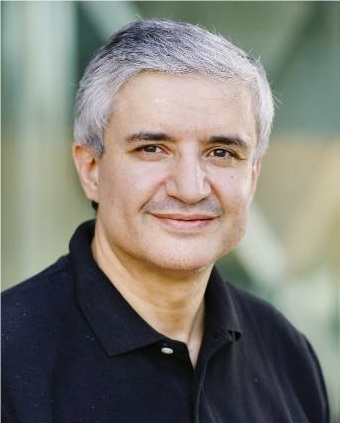}}]{Mohamed-Slim Alouini}
(S'94-M'98-SM'03-F'09)  was born in Tunis, Tunisia. He received the Ph.D. degree in Electrical Engineering
from the California Institute of Technology (Caltech), Pasadena,
CA, USA, in 1998. He served as a faculty member in the University of Minnesota,
Minneapolis, MN, USA, then in the Texas A\&M University at Qatar,
Education City, Doha, Qatar before joining King Abdullah University of
Science and Technology (KAUST), Thuwal, Makkah Province, Saudi
Arabia as a Professor of Electrical Engineering in 2009. His current
research interests include the modeling, design, and
performance analysis of wireless communication systems.
\end{IEEEbiography}
\end{document}